\documentclass[aps, prd, twocolumn, floatfix, nofootinbib, superscriptaddress]{revtex4-1}
\usepackage{amsmath,amsfonts,amssymb,bm}
\usepackage{dsfont}
\usepackage{graphicx}

\usepackage{epstopdf}
\usepackage{color}
\usepackage{soul}
\usepackage{slashed}
\usepackage{pifont}
\usepackage[mathscr,scaled=1.15]{urwchancal}
\DeclareFontFamily{OT1}{pzc}{}
\DeclareFontShape{OT1}{pzc}{m}{it}%
{<-> s * [1.15] pzcmi7t}{}
\DeclareMathAlphabet{\mathpzc}{OT1}{pzc}{m}{it}

\definecolor{purple}{rgb}{0.5,0,0.5}
\definecolor{blue}{rgb}{0.0,0,0.9}
\definecolor{prdblue}{rgb}{0.133,0.118,0.498}
\usepackage[colorlinks=true, pdfstartview=FitV, linkcolor=prdblue, citecolor= prdblue, urlcolor=prdblue]{hyperref}

\newcommand{\lan}{\left(}
\newcommand{\ran}{\right)}
\newcommand{\lsan}{\left[}
\newcommand{\rsan}{\right]}

\begin{document}

\title{Dressed Quark Tensor Vertex and Nucleon Tensor Charge}

\author{Langtian Liu}
\affiliation{Department of Physics and State Key Laboratory of Nuclear Physics and Technology, Peking University, Beijing 100871, China}
\author{Lei Chang}
\affiliation{School of Physics, Nankai University, Tianjin 300071, China}
\author{Yu-xin Liu}
\affiliation{Department of Physics and State Key Laboratory of Nuclear Physics and Technology, Peking University, Beijing 100871, China}
\affiliation{Collaborative Innovation Center of Quantum Matter, Beijing 100871, China}
\affiliation{Center for High Energy Physics, Peking University, Beijing 100871, China}

%\date{21 Feb 2018}

\begin{abstract}
We construct the quark-antiquark scattering kernels of Bethe-Salpeter equation from the quark self-energy directly under two specific forms of quark-gluon vertices. The quark dressed tensor vertex is then calculated within  this consistent framework and rainbow-ladder(RL) approximation.  After employing a simplified nucleon model,  the nucleon tensor charge can be defined with the  tensor vertex. We then compute the tensor charge with the bare tensor vertex and the dressed vertices obtained in this framework and in RL approximation. The obtained results are consistent with the lattice QCD calculations. We also find  that typically the  gluon dressing effects suppress the nucleon tensor charge compared to the bare tensor vertex,  by about $23\%$ for RL approximation, and  turn to be about $13\%$ in this framework.
\end{abstract}

\maketitle
%%%%%%%%%%%%%%%%%%%%%%%%%%%%%%%%%%%%%%%%%%%%%%%%%%%%%%%%%%%%%%%%%%%%%%%%%%%%%%%%%%%%
\section{Introduction}
%%%%%%%%%%%%%%%%%%%%%%%%%%%%%%%%%%%%%%%%%%%%%%%%%%%%%%%%%%%%%%%%%%%%%%%%%%%%%%%%%%%%%%%

Parton distributions  play an essential role in high-energy nuclear and particle physics, and today there are vast international experimental programs aimed at their measurements. At leading twist level, there are three parton distributions to describe the inner structure of hadrons: the number density $f(x)$; the longitudinal polarization distribution $g(x)$; the transversity  distribution $h(x)$ \cite{Barone:2001sp}.
The transversity distribution $h(x)$ describes the net quark density of  transversally polarized quark in the transversally polarized hadrons, i.e., the number density of quark with momentum fraction $x$ and spin parallel to the hadron transverse polarization direction minus the density of quark with the same momentum fraction but antiparallel.  The tensor charge can be read from the first moment of transversity  distribution $h(x)$  via \cite{PhysRevD.52.2960,PhysRevLett.67.552,JAFFE1992527}:
\begin{equation}
\delta q = \int_{-1}^1 dx h(x) = \int_0^1 dx \big[h(x) - \bar h(x)\big]\, ,
\end{equation}
where $\bar h(x)$  denotes to the transversity distribution of  antiquark in the hadrons.The relation $h(-x) = -\bar h(x), \; x > 0$ is employed in the second step.  In matrix elements language, the tensor charge can be extracted from nucleons as
\begin{equation}
\delta q\; \bar u(p, s) \sigma^{\mu \nu} u(p, s) = \langle p,s|\bar \psi_q \sigma^{\mu \nu} \psi_q|p,s\rangle\, .
\end{equation}
Here, $|p, s\rangle$ is the nucleon state vector with momentum $p$ and spin $s$ and $u(p ,s)$ is nucleon spinor.  

The tensor charge can be related to the nucleon electric dipole moment (EDM) which is sensitive to the additional $\mathcal{CP}$ violation so that we can use it to search for new physics beyond the standard model \cite{PhysRevD.97.074018,PhysRevD.88.074036,Pitschmann:2014jxa}. So there are some works about the nucleon tensor charges up to now both experimentally and theoretically.

Recently the extraction of the quark transversity distributions from experimental data became possible at some accessible scales \cite{PhysRevD.87.094019,Ye:2016prn,PhysRevLett.120.192001}. Noticing the tensor charge is scale dependent,  the scale dependence can be obtained via  perturbative QCD theory as \cite{Barone:2001sp,Barone:1997fh}:
\begin{equation}
\delta q (Q^2) = \delta q(Q_0^2) \Big[\frac{\alpha_s(Q_0^2)}{\alpha_s(Q^2)}\Big]^{-4/(33-2n_f)}\, ,
\end{equation}
where $\alpha_s(k^2)$ is the running coupling constant of strong interaction and $n_f$ is the number of flavor. Hereinafter we choose $n_f = 4$.

The extraction of nucleon tensor charges from global analysis of electron-proton and proton-proton data gives \cite{PhysRevLett.120.192001}:
\begin{align*}
& \delta u = 0.39 \pm 0.10\, ,\\
& \delta d = -0.11 \pm 0.26\, ,
\end{align*}
at $Q^2 = 4 \text{GeV}^2$.
All these experimental data analysis indicate suppression compared with the constituent quark model predictions, which are  $\delta u=\frac{4}{3}$ and $\delta d=-\frac{1}{3}$ in the nonrelativistic limit \cite{PhysRevD.97.074018}.

The suppression is  also confirmed by lattice QCD (lQCD) simulations as well as some phenomenological approaches \cite{PhysRevD.94.054508,Alexandrou:2017qyt,Gupta:2018lvp}.  In Ref.\cite{Alexandrou:2017qyt}, the authors calculated the nucleon tensor charges within lQCD using simulations generated with two dynamical degenerate light quarks with masses fixed to reproduce approximately the physical pion mass, and they give:
\begin{align*}
& \delta u = 0.782(16)(2)(13)\,,\\
& \delta d = -0.219(10)(2)(13)\,,
\end{align*}
at $Q^2 = 4 \text{GeV}^2$. The results are also much smaller compared to the constituent quark model predictions,  although the results from lQCD and experiments are quantitatively  inconsistent at present.

In this paper, we compute the quark tensor vertex in a consistent way via Dyson-Schwinger equation (DSE) and Bethe-Salpeter equation (BSE) and  obtain the tensor charges by inserting the tensor vertex into a simplified nucleon model --- quark + scalar diquark \cite{Hobbs:2016xfz}. We directly  derive the two-particle irreducible scattering kernel from the quark self-energy  and find it is consistent with the  constructed that scattering kernel from symmetry constraints \cite{Chang:2009zb}. We also find that the dressed effects are very important  to the nucleon tensor charge. The rainbow-ladder processes suppresses the tensor charge by about $23\%$  comparing to that in tree-level tensor vertex. Furthermore, when we go beyond the rainbow-ladder truncation scheme  with the quark-gluon interaction vertex in Munczek model \cite{Munczek:1994zz}, which satisfies Ward-Takahashi identity, the tensor charges grow up a bit and the suppression of tensor charge  turns to be $13\%$. Our result is consistent with that from lQCD simulations.

The remainder of this paper is organized as follows: In Sec.\ref{sec:ten-fomu} we construct the BSE for the general vertex amplitude when we choose a quark-gluon vertex beyond rainbow-ladder approximation; In Sec.\ref{sec:sol} we give out the solution of quark and quark tensor vertex with Munczek vertex, and as a comparison, we compare with the solution in rainbow-ladder approximation at the same quark condensates. In Sec.\ref{sec:ten-char} we substitute the quark tensor vertex into  a simplified nucleon model to estimate the dress effects on tensor charge. Finally, we summarize in Sec.\ref{sec:sum}.

%%%%%%%%%%%%%%%%%%%%%%%%%%%%%%%%%%%%%%%%%%%%%%%%%%%%%%%%%%%%%%%%%%%%%%%%%%%%%%
\section{Constructing Tensor Vertex Equation}
\label{sec:ten-fomu}
%%%%%%%%%%%%%%%%%%%%%%%%%%%%%%%%%%%%%%%%%%%%%%%%%%%%%%%%%%%%%%%%%%%%%%%%%%%%%%

Let's first begin with the general form of the quark Dyson-Schwinger equation,
\begin{equation}
\label{ds-q-re}
\begin{split}
S^{-1}(p)= Z_2(\mu, \Lambda) S_{0}^{-1}(p)+\Sigma(p) ,
\end{split}
\end{equation}
with
\begin{equation}
 \Sigma(p)=Z_2(\mu, \Lambda) g^2\int_{dq}^{\Lambda}D_{\alpha \beta}(k)t^a\gamma_\alpha S(q) t^a\Gamma_{\beta}(q,p),
\end{equation}
where $S,D$ are the dressed-quark and -gluon propagators respectively. $\Gamma_{\mu}(q,p)$ is the general quark-gluon vertex. $S_{0}^{-1}(p)=i\gamma\cdot p+Z_{m}(\mu, \Lambda) m(\mu)$ is the quark propagator at renormalization point $\mu$. $Z_2(\mu, \Lambda), Z_m(\mu, \Lambda)$ are quark field strength renormalization coefficient and quark mass renomalization coefficient respectively and we denote $Z_4(\mu, \Lambda) = Z_2(\mu, \Lambda)*Z_m(\mu, \Lambda)$. $\int_{dq}^{\Lambda}$ represents a Poincar$\acute{e}$ invariant regularization of the four-dimensional integral, with $\Lambda$ the regularization mass scale.  $k = q-p$ is the momentum carried by gluon.

In practical calculation, we parameterize the gluon propagator in Landau gauge.
\begin{equation}
g^2 D_{\alpha \beta}(k) = \mathcal{G}(k^2)(\delta_{\alpha \beta} - \frac{k_{\alpha} k_{\beta}}{k^2})\, ,
\end{equation}
where
\begin{align}
&\mathcal{G}(k^2) =\mathcal{G}_{ir}(k^2) + \mathcal{G}_{uv}(k^2)\,,\\
&\mathcal{G}_{ir}(k^2) = \frac{8 \pi^2}{\omega^5} m_{g}^{3} e^{-k^2/\omega^2}\,,\\
&\mathcal{G}_{uv}(k^2) = \frac{8 \pi^2 \gamma_m}{\ln[\tau+(1+k^2/\Lambda_{\text{QCD}}^2)^2]} \frac{1-e^{-k^2/4m_t^2}}{k^2}\, .
\end{align}
$\omega, m_{g}$ are the interaction parameters of infrared gluon propagator and they have dimension $[\text{GeV}]$, denoting the interaction width and magnitude respectively.
In the ultraviolet gluon propagator, $\gamma_m = 12/(33-2n_f)$ is the anomalous dimension and $n_f$ is the number of flavors. In our calculations, we choose $n_f = 4$. $\Lambda_{\text{QCD}}$ is the QCD scale that characterize the non-perturbative QCD and we choose $\Lambda_{\text{QCD}} = 0.234\text{GeV}$. $\tau = e^2 - 1, m_t = 0.5\text{GeV}$. Here $e$ is the natural constant.

The quark propagator can be parameterized as
\begin{equation}\label{ABf}
S^{-1}(p) = i \gamma \cdot p A(p^2) + B(p^2)\, .
\end{equation}
Usually, the loop integral in quark self-energy will be UV divergent in four dimensional space-time. We choose a quark current mass independent renomalization scheme in  chiral limit to renormalize the quark solution:
\begin{align}
A(p^2)&\big|_{p^2 = \mu^2,\text{chiral limit}} = 1\,,\\
\frac{\partial B(p^2)}{\partial m(\mu)}&\big|_{p^2 = \mu^2, \text{chiral limit}} = 1\, ,
\end{align}
where the chiral limit condition is defined unambiguously by
$Z_{4}m(\mu)|_{\Lambda\to\infty}=0$ in QCD.

We then address the Munczek ansatz for  quark-gluon vertex\cite{Munczek:1994zz}:
\begin{equation}
\label{vertex}
i \Gamma_\nu^{a}(p+k,p) =t^{a} \frac{\partial}{\partial p^\nu} \int_0^1 d \alpha S^{-1}(p+\alpha k) + i \Gamma_\nu^{aT}(p+k,p)\, ,
\end{equation}
where $\Gamma_\nu^{aT}(p+k,p)$ is the transverse part of the quark-gluon vertex which satisfies $k^\nu \Gamma_\nu^{aT}(p+k,p) = 0$. At the first stage, we will use the longitudinal part of the quark-gluon vertex, which  is an analytical form and therefore make it possible to derive the two-particle irreducible scattering kernel. For the consistent extension of modeling transverse part, one might track to Ref.\cite{Qin:2013mta}.

Let's denote a dressed quark vertexes as $\Gamma(k; 0)$ regardless of the Lorentz indices. The general form of Bethe-Salpeter equation for the dressed quark vertex can be written as
\begin{equation}
\label{eq-bs}
\begin{split}
[\Gamma(k;0)]_{EF} &= Z_v [\Gamma_0(k;0)]_{EF} \\
&\quad + \int_{dq}^\Lambda [K(k,q;0)]^{GH}_{EF}[\chi(q;0)]_{GH}\, .
\end{split}
\end{equation}
Here, $Z_v$ is the renormalization coefficient for the dressed quark vertex and inhomogeneous term $\Gamma_0(k;0)$ is the tree level term of  the dressed quark vertex. $K(k, q; 0)$ is the two-particle irreducible scattering kernel. $\chi(q;0)=S(q)\Gamma(q;0)S(q)$ denotes the unamputated dressed quark vertex and $E,F,G,H$ are spinor indices here.

As described in Ref.\cite{Munczek:1994zz, Cornwall:1974vz,McKay:1989rk}, for the colorless bilocal field, which represent the colorless vertexes  (such as quark-photon vertex and the tensor vertex considered below), we can derive the two-particle irreducible scattering kernel from the composite operator effective action:
\begin{equation}
[K(k,q;0)]^{GH}_{EF} = - \frac{\delta [\Sigma(k)]_{EF}}{\delta [S(q)]_{GH}}\, ,
\label{eq-scaker}
\end{equation}
which fulfills the consistency between the DSE and BSE. The quark self energy which had been given in Eq.\eqref{ds-q-re} can be written explicitly as:
\begin{equation}
\begin{split}
[\Sigma(k)]_{EF} &= Z_2 g^2\int_{dl}^\Lambda D_{\mu \nu}(l-k)\\
&\quad\times t^a[\gamma_\mu]_{EM}[S(l)]_{MN}t^a[\Gamma_\nu(l,k)]_{NF} \, .
\end{split}
\end{equation}
The subscripts $M, N$ also denote the spinor indices.

 Taking an approximation that the gluon propagator does not depend on quark propagator, i.e, $\delta D_{\mu \nu}/\delta S = 0$  , and considering the dependence of the quark-gluon vertex on $S$, we have
\begin{equation}
[K(k,q;0)]^{GH}_{EF} = - \frac{\delta [\Sigma(k)]_{EF}}{\delta [S(q)]_{GH}} = \text{\ding{172}} + \text{\ding{173}}\, ,
\end{equation}
with

\begin{subequations}
\begin{align}
&\text{\ding{172}}=-Z_2g^2D_{\mu \nu}(q-k)t^a\left[\gamma_\mu\right]_{EG}t^a\left[\Gamma_\nu(q,k)\right]_{HF}\, ,\\
\nonumber&\text{\ding{173}}=-Z_2g^2\int_{dl}^\Lambda D_{\mu \nu}(l-k)\\
&\quad\quad\times t^a\left[\gamma_\mu\right]_{EM}\left[S(l)\right]_{MN}t^a \frac{\delta \left[\Gamma_\nu(l,k)\right]_{NF}}{\delta \left[S(q)\right]_{GH}}\, .
\end{align}
\end{subequations}
After doing  functional derivation  on this vertex with respect to quark propagator, we obtain
\begin{equation}
\begin{split}
&\text{\ding{173}}=Z_2g^2\int_{dl}^\Lambda D_{\mu \nu}(l-k) t^a \lsan \gamma_\mu\rsan_{EM}\lsan S(l)\rsan_{MN}t^a\\
&\qquad\times \frac{\partial}{i\partial k^\nu}\int_0^1d \alpha \lsan S^{-1}(k+\alpha(l-k))\rsan_{NG}\\
&\qquad\times\delta^{(4)}\lan k+\alpha(l-k)-q\ran\lsan S^{-1}\lan k+\alpha(l-k)\ran\rsan_{HF}\,.
\end{split}
\end{equation}
Substituting this  into general BS equation (Eq.\eqref{eq-bs}), we get:

\begin{widetext}
\begin{equation}
\begin{split}
\lsan \Gamma(k;0) \rsan_{EF} &= Z_v [\Gamma_0(k;0)]_{EF} - Z_2g^2\int_{dq}^\Lambda D_{\mu \nu}(q-k)t^a\lsan \gamma_\mu \rsan_{EG}\lsan \chi(q;0) \rsan_{GH}t^a\lsan \Gamma_\nu(q,k)\rsan_{HF}\\
&\quad+Z_2g^2\int_{dq}^\Lambda D_{\mu \nu}(q-k)t^a\lsan \gamma_\mu\rsan_{EM}\lsan S(q) \rsan_{MN}t^a\left[\Lambda_{\nu}(q,k;0)\right]_{NF}\, ,
\end{split}
\end{equation}
\end{widetext}

with
\begin{equation}
\Lambda_\nu(q,k;0) = \frac{\partial}{i\partial k^\nu}\int_0^1 d \alpha \Gamma(k+\alpha(q-k);0)\, .
\end{equation}
It's easy to see that this term satisfies
\begin{equation}
\label{eq-secker-id}
i q_\beta \Lambda_\beta(q+k,k;0) = \Gamma(q+k;0) - \Gamma(k;0)\,.
\end{equation}
The relation in Eq.\eqref{eq-secker-id} is  consistent with the constraint obtained from the symmetries --- vector or axial vector Ward identities \cite{Chang:2009zb},  where the constraint for $\Lambda_{5\beta}$ of pseudoscalar mesons was given with the axial vector Ward identity and realized practically in the way similar to getting the Ball-Chiu quark-gluon vertex \cite{Ball:1980ay,Ball:1980ax}. Here, we derive the $\Lambda_\beta$ directly for general BS amplitude even the corresponding amplitude doesn't have a constraint from symmetry, e.g., tensor vertex we discuss in this paper.   It's then clear that this direct way from quark self energy and indirect way from symmetries are equivalent and the direct way is more powerful.

\begin{figure}[b]
\centering
\includegraphics[width=0.45\textwidth]{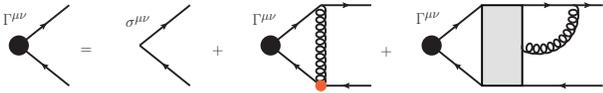}
\caption{The Feynman diagram of tensor vertex BS equation.}
\label{fig-ten-bs}
\end{figure}

As for quark tensor vertex, let's denote
\begin{equation}
\Lambda^{\mu \nu}_\beta(q,k;0) = \frac{\partial}{i\partial k^\beta}\int_0^1 d \alpha \Gamma^{\mu \nu}(k+\alpha(q-k);0)\,.
\end{equation}
The BSE for quark tensor vertex amplitude can be written as
\begin{eqnarray}
&&\Gamma^{\mu \nu}(p;0) = Z_t \sigma^{\mu \nu} \nonumber\\
&&\qquad - \frac{4}{3}Z_2 g^2\int_{dq}^{\Lambda}D_{\alpha \beta}(k) \gamma_\alpha S(q) \Gamma^{\mu \nu}(q; 0) S(q) \Gamma_\beta(q,p) \nonumber\\
&&\qquad +\frac{4}{3}Z_2 g^2\int_{dq}^{\Lambda}D_{\alpha \beta}(k)\gamma_\alpha S(q) \Lambda^{\mu \nu}_\beta(q,p;0)\,.
\end{eqnarray}
The coefficient $4/3$ comes from the color factor $t^a t^a = 4/3$ and the schematic diagram of tensor vertex BSE is showed in Fig.~\ref{fig-ten-bs}.
The $Z_t$ is the quark tensor vertex renormalization coefficient and determined  by requiring $\Gamma^{\mu \nu}(p;0)\big|_{p^2=\mu^2} = \sigma^{\mu \nu}$ at the renormalization point $p^2=\mu^2$ in practical computation.
If parameterizing the general form of the quark tensor vertex as
\begin{eqnarray}
\Gamma_{\mu \nu}(p; 0) &&= \sigma_{\mu \nu} E(p^2) + \big((i \gamma \cdot p) \sigma_{\mu \nu}+\sigma_{\mu \nu} (i \gamma \cdot p)\big) F(p^2) \nonumber\\
&&\quad+\big((i \gamma \cdot p) \sigma_{\mu \nu}-\sigma_{\mu \nu} (i \gamma \cdot p)\big) G(p^2) \nonumber\\
&&\quad+ (i \gamma \cdot p) \sigma_{\mu \nu} (i \gamma \cdot p) H(p^2)\,,\nonumber
\label{tenver-decompose}
\end{eqnarray}
the renormalization procedure is simply the requirement $E(p^{2}=\mu^2) = 1$.
%%%%%%%%%%%%%%%%%%%%%%%%%%%%%%%%%%%%%%%%%%%%%%%%%%
\section{Properties of Quark Tensor Vertex}
\label{sec:sol}
%%%%%%%%%%%%%%%%%%%%%%%%%%%%%%%%%%%%%%%%%%%%%%%%%%

Eq. (\ref{vertex})  can be expanded  with Eq.(\ref{ABf}) as:
\begin{equation}
\begin{split}
&\Gamma_\nu(p+k,p) \\
&\quad= \frac{\partial}{i \partial p^\nu}\int_0^1 d \alpha S^{-1}(p+\alpha k)\\
&\quad= \int_0^1 d \alpha \big[\gamma_\nu A((p+\alpha k)^2)\\
&\quad\qquad\quad+2 (p_\nu+\alpha k_\nu) \gamma \cdot (p+\alpha k)A'((p+\alpha k)^2)\\
&\qquad~~~~~~-2i(p_\nu+\alpha k_\nu)B'((p+\alpha k)^2)\big]\, ,
\end{split}
\end{equation}
where $k = q - p$. By projection of the quark-DSE Eq.\eqref{ds-q-re}, we obtain two coupled differential-integral equations:
\begin{subequations}
\begin{align}
&\nonumber A(p^2) = Z_2+Z_2 \frac{4}{3}\int_{dq}^{\Lambda} \mathcal{G}(k)\int_0^1 d \alpha \Big[ c_{aa}A(q^2) A((p+\alpha k)^2)\\
\nonumber&~~~~+c_{ada}A(q^2)A'((p+\alpha k)^2) + c_{bdb}B(q^2)B'((p+\alpha k)^2)\Big] \\
&\qquad\Big/\Big(q^2 A^2(q^2)+B^2(q^2)\Big)\, ,\\
&\nonumber B(p^2) = Z_4 m + Z_2 \frac{4}{3}\int_{dq}^{\Lambda} \mathcal{G}(k)\int_0^1 d \alpha \Big[c_{ba}B(q^2)A((p+\alpha k)^2)\\
\nonumber&~~~~+c_{bda}B(q^2)A'((p+\alpha k)^2)+c_{adb}A(q^2)B((p+\alpha k)^2)\Big] \\
&\qquad \Big/\Big(q^2 A^2(q^2) + B^2(q^2)\Big)\, .
\end{align}
\end{subequations}
with $\phi'(p^2) = \partial \phi(p^2) / \partial p^2$.
The coefficients are listed below:
\begin{align}
&c_{aa} = \frac{p \cdot q}{p^2}+2 \frac{(p \cdot k)(q \cdot k)}{p^2 k^2}\, ,\\
&c_{ada} = -2\alpha (k \cdot q) + \frac{\alpha(k \cdot p) + p^2}{p^2}\times\nonumber\\
&\qquad\quad \big[2\frac{(k \cdot p)(k \cdot q)}{k^2}-2(p\cdot q)\big]\, ,\\
&c_{bdb} = 2 (\frac{(k \cdot p)^2}{k^2 p^2} - 1)\, ,\\
&c_{ba} = 3\, ,\\
&c_{bda} = 2 (p^2 - \frac{(k \cdot p)^2}{k^2})\, ,\\
&c_{adb} = -2 (p \cdot q - \frac{(k \cdot q)(k \cdot p)}{k^2})\, .
\end{align}

The rainbow-ladder truncation of Dyson-Schwinger equations and Bethe-Salpeter equations has been implemented successfully in hadron physics, especially for the description of pseudoscalar and vector channels, see for example Refs \cite{Maris:2003vk,Roberts:2007ji,Qin:2011dd,Qin:2011xq,Bashir:2012fs,Cloet:2013jya}. The interaction parameter within the truncation has been fitted by reproducing ground pseudoscalar meson properties. When one goes beyond rainbow-ladder truncation one should tune the parameter to match the same observable quantities. However it is beyond our interest now to get the on shell meson properties. To compare the results between these two different truncations we enforce the chiral quark condensate taking the same value which is supposed to characterize the magnitude of dynamical chiral symmetry breaking (DCSB). 

The quark condensate in chiral limit, $\langle 0|\bar \psi \psi |0\rangle$  is calculated in the general way with the quark propagator \cite{Roberts:1994dr}:
\begin{equation}
\label{con}
\begin{split}
\langle 0|\bar \psi \psi |0\rangle &= - \text{Tr}[S(p)]\\
&= -3 \int \frac{d^4p}{(2\pi)^4}\text{Tr}\left[\frac{-i \gamma \cdot p A(p^2) + B(p^2)}{p^2 A^2(p^2) + B^2(p^2)}\right]\\
&= -\frac{3}{(2\pi)^2}\int_0^\infty d p^2 \frac{p^2 B(p^2)}{p^2 A^2(p^2) + B^2(p^2)}\,.
\end{split}
\end{equation}
Eq.~\eqref{con} is the unrenormalized quark condensate and will diverge logarithmically. The renormalisation factor for the condensate can be derived from the  vector Ward-Takahashi identity whose two legs are with unequal masses , that is $Z_4 = Z_2*Z_m$. We can then renormalize the quark condensate by multiplying $Z_4$ on the r.h.s of Eq.\eqref{con} and then set $\Lambda\to\infty$ at final, i.e., $\int_0^\infty d p^2\to \lim\limits_{\Lambda\to\infty}Z_{4}(\mu,\Lambda)\int_{0}^{\Lambda^{2}}d p^{2}$.

Here we set the parameters to get the quark condensate $\langle\bar\psi\psi\rangle = -(0.225\text{GeV})^3$ at the renormalization point $\mu = 2\text{GeV}$. The parameters of gluon propagator in these two scheme are showed in Tab.~\ref{tab-g-para}.  When fitting the parameters,  we not only require the quark condensate is the same as the case in rainbow ladder approximation, but also set $\omega$ to make the quark condensate at the minimum point as a function of $\omega$ in order to get the maximal chiral symmetry breaking vacuum.
\begin{figure}
[t]
\includegraphics[width=0.5\textwidth]{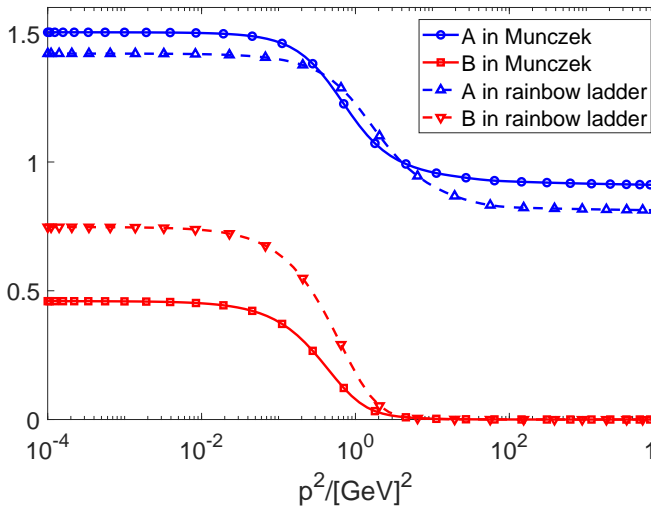}
\caption{(Color online)The quark solutions with the same quark condensate $\langle\bar\psi\psi\rangle = -(0.225\text{GeV})^3$ at the renormalization point $\mu = 2\text{GeV}$.}
\label{quark-com}
\end{figure}

\begin{table}[b]
\centering
\caption{The gluon parameters for rainbow-ladder and beyond rainbow-ladder truncation scheme. The quark solutions have the same quark condensates $\langle\bar\psi\psi\rangle = -(0.225\text{GeV})^3$ at the renormalization point $\mu = 2\text{GeV}$.}
\begin{tabular}{|c|c|c|}
\hline
truncation scheme  &  $m_{g}/[\text{GeV}]$   & $\omega/[\text{GeV}]$\\
\hline
rainbow-ladder &  $0.82$   &  0.5\\
\hline
Munczek   & 0.436  &  0.355\\
\hline
\end{tabular}
\label{tab-g-para}
\end{table}

 The solutions of quark propagator are shown  in Fig.\ref{quark-com} compared to the solution in RL approximation.
Although the two ansatz  of quark-gluon vertex lead to  the same quark condensate --- the same magnitude of dynamical chiral symmetry breaking, the strength of quark-gluon interaction in Munczek ansatz is much weaker than that in the RL approximation, quantitatively, about $53\%$, which is comparable with the physical coupling strength of QCD\cite{BINOSI2015183}.

\begin{figure}
[h]
\includegraphics[width=0.5\textwidth]{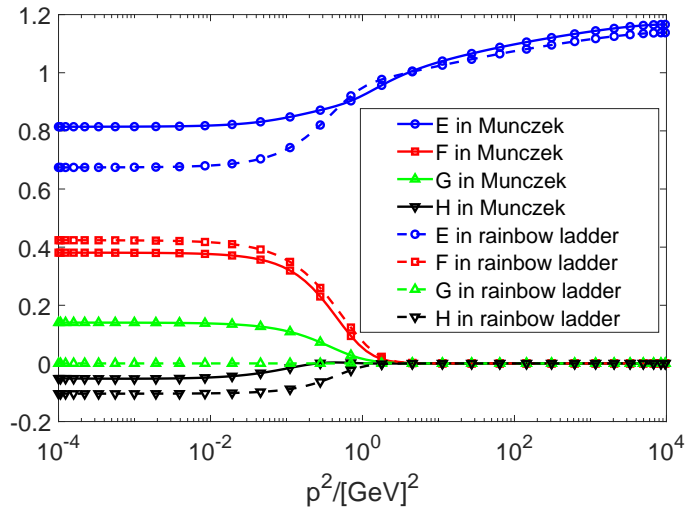}
\caption{(Color online)The comparison of quark tensor vertex under the same quark condensate $\langle\bar\psi\psi\rangle = -(0.225\text{GeV})^3$ at renormalization point $\mu = 2\text{GeV}$.}
\label{tencharcompare}
\end{figure}

With the quark tensor vertex decomposition form Eq.\eqref{tenver-decompose} and by projection, we can get the differential-integral equations  for tensor vertex. In RL approximation, the scalar function $G(p^2)$  decouples from other scalar functions and  will always be zero \cite{PhysRevD.88.074036}.  When  going beyond the rainbow-ladder approximation,  all four scalar functions in Eq.\eqref{tenver-decompose} are coupled and one needs to solve the whole coupled differential-integral equations. By the Chebyshev interpolation method \cite{Bloch:1995dd}, we can solve the quark tensor vertex amplitude numerically. The solutions of quark tensor vertex in RL approximation and with Munczek ansatz are showed in Fig.\ref{tencharcompare}.

We can see that the  scalar functions $E(p^2), F(p^2), G(p^2), H(p^2)$ in RL approximation and Munczek ansatz  have large differences in the infrared region. With this quark tensor vertex,  we can then compute the nucleon tensor charge.
%%%%%%%%%%%%%%%%%%%%%%%%%%%%%%%%%%%%%%%%%%%%%%%%%%%%%%%%%%%%%%%%%%%%%%%%%%%%%%%%%%%%%%
\section{Dress Effects on The Tensor Charge}
\label{sec:ten-char}
%%%%%%%%%%%%%%%%%%%%%%%%%%%%%%%%%%%%%%%%%%%%%%%%%%%%%%%%%%%%%%%%%%%%%%%%%%%%%%%%%%%%%
In principle, to calculate the nucleon tensor charge, we need to know the information of the nucleon wave functions  by solving the Faddeev equation, see for example Ref.\cite{PhysRevD.98.054019}.
Our aim is to seek the gluon dressing effect on nucleon tensor charge and it would be difficult and expensive for us to operate this calculation if  solving the coupled gap equation, vertex equation and Faddeev equation. Thus to achieve this goal  we  make a simplified ansatz for nucleon structure and test the dressing effects in a  simplified way.

Recalling Ref. \cite{Hobbs:2016xfz}, the authors built a phenomenological picture of nucleon and studied the dressing effects on axial charge. The nucleon contains the constituent quark and point-like scalar diquark, and   the quark-diquark-nucleon interaction amplitude can be characterized as
\begin{equation}
\phi(k^2) = g \frac{\Lambda^2}{k^2 + \Lambda^2}\, ,
\end{equation}
with $k$ the momentum carried by quark.
The schematic diagram of photon coupled to the quark is depicted in Fig.\ref{fig-proton-vec} which can be expressed explicitly as:
\begin{figure}[h]
\centering
\includegraphics[width=0.4\textwidth]{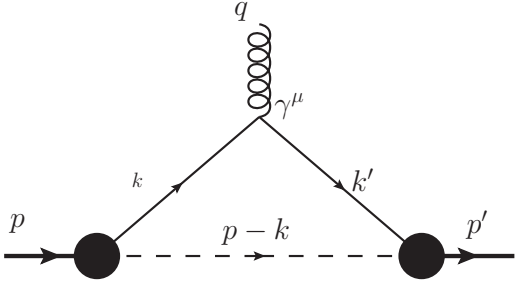}
\caption{The first nontrival contribution of vector vertex in the nucleon. In the diagram, the internal solid lines represent for quarks, while dashed lines are the scalar diquarks. The filled circles are the momentum dependent quark-diquark-nucleon interaction $\phi(k^2)$.}
\label{fig-proton-vec}
\end{figure}
\begin{equation}
\begin{split}
&\bar u(p') \Gamma^\mu(p', p) u(p) \\
&= \int \frac{d^4k}{(2\pi)^4}\frac{1}{i\slashed{k}\prime+m}\gamma^{\mu}\frac{1}{i\slashed{k}+m}\frac{\phi(k^{\prime 2}) \phi(k^2)}{(p-k)^2+m_D^2}\, .
\end{split}
\label{eq-em}
\end{equation}
From Eq.\eqref{eq-em}, one can calculate the electromagnetic form factors. By fitting to the experimental data, one can fix the model parameter: the quark constituent mass $m$, the scalar diquark mass $m_D$, the quark-diquark-nucleon interaction strength $g$ and width $\Lambda$. The parameters adapted from \cite{Hobbs:2016xfz} are showed in Table.\ref{tab-nucl-para}
\begin{table}[b]
\centering
\caption{The parameters come from fitting to the electromagnetic form factors \cite{Hobbs:2016xfz}. There they consider the contribution of bare vector vertex.}
\begin{tabular}{|c|c|c|c|}
\hline
$m/[\text{GeV}]$     &  $m_D/[\text{GeV}]$    &   $\Lambda/[\text{GeV}]$    & $g$    \\
\hline
0.637    &  0.947     &     0.228       & 79.104   \\
\hline
\end{tabular}
\label{tab-nucl-para}
\end{table}

Similarly, we can couple the quark tensor vertex into this nucleon model to calculate the tensor charge of $u$ quark.
We get the $u$ quark tensor charge in this simplified proton model as
\begin{equation}
\begin{split}
&\delta q\; \bar u(p, s) \sigma^{\mu \nu} u(p, s) \\
&= \langle p,s|\bar \psi_q \sigma^{\mu \nu} \psi_q|p,s\rangle\\
&=\int \frac{d^4k}{(2\pi)^4}\frac{1}{i\slashed{k}+m}\Gamma^{\mu \nu}(k;0)\frac{1}{i\slashed{k}+m}\frac{|\phi(k^2)|^2}{(p-k)^2+m_D^2}\, ,
\end{split}
\label{eq-ten}
\end{equation}
where $p, s$ are the proton momentum and spin respectively. $u(p, s)$ is the proton spinor and $\delta q$ is the tensor charge with
\begin{equation}
\delta q = \int dk^2 h(k^2)\, .
\end{equation}
Here we will calculate the proton $u$ quark tensor charge in three type of quark tensor vertex: bare tensor vertex, dressed quark tensor vertex in RL approximation and the one with Munczek ansatz.

\begin{figure}
[t]
\centering
\includegraphics[width=0.5\textwidth]{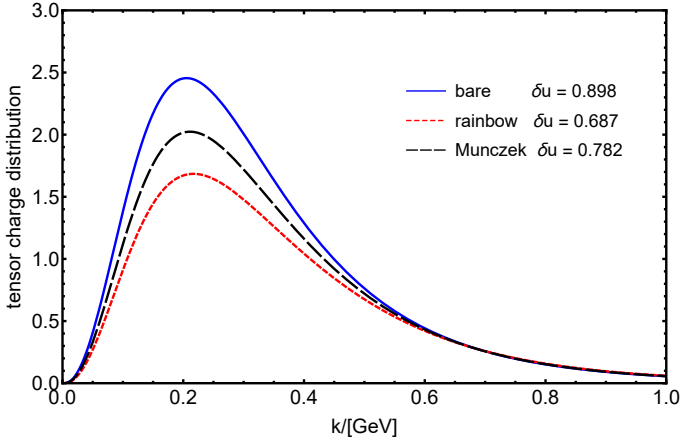}
\caption{A comparison for tensor charge distribution functions of $u$ quark in proton under bare tensor vertex, dressed tensor vertex in rainbow-ladder approximation and beyond rainbow-ladder approximation. Here we plot the tensor charge distribution function as $2 k\, h(k^2)$.}
\label{fig-ten-dist}
\end{figure}

For the case of bare tensor vertex $\Gamma^{\mu \nu}(k;0) = \sigma^{\mu \nu}$ following the calculation method in \cite{Hobbs:2016xfz} (the crucial step is that the authors consider the Gegenbauer polynomial expansion of diquark propagator and then the angular part  $k\cdot p$ pieces can be integrated out analytically by the orthogonality. The momentum existed in the charge distributions is the one carried by quark), we arrive at
\begin{equation}
\begin{split}
h_{bare}(k^2) &= (\frac{g \Lambda^2}{4 \pi})^2 \frac{Z k^2}{(k^2 + m^2)^2 (k^2 + \Lambda^2)^2}\times\\
&\quad \big[m^2 + M^2 \frac{(Z k^2)^2}{3} + m M Z k^2\big]\, ,
\end{split}
\label{ten-dist-bare}
\end{equation}
here $M$ is the proton mass. We adopt $M = 0.938 \text{GeV}$ from PDG data and
\begin{equation}
\begin{split}
Z &= \frac{-1}{2 M^2 k^2} \Big(k^2 + m_D^2 \\
&\quad- M^2 - \sqrt{(k^2 + m_D^2 - M^2)^2 + 4 M^2 k^2}\Big)\, .
\end{split}
\end{equation}
To arrive at the tensor charge distribution Eq.\eqref{ten-dist-bare} under tree level tensor vertex, we need to make use of the spinor technology in Euclidean space.
\begin{equation}
u(p, s) \bar u(p, s) = \frac{1}{2}(1 + i \gamma_5 \slashed s)(i \slashed p - M)\, ,
\end{equation}
$p, M, s$ are the momentum vector, mass, and spin vector of the fermion which the spinor $u(p, s)$ describes, respectively. Then we get a  formula to compute the quantity $\bar u(p, s) \Gamma u(p, s)$ where $\Gamma$ is an arbitrary Dirac matrix or their combination:
\begin{equation}
\bar u(p, s) \Gamma u(p, s) = \text{Tr}\Big[\frac{1}{2}(1 + i \gamma_5 \slashed s)(i \slashed p - M)  \Gamma\Big]\, .
\end{equation}
We show this tensor charge distribution of $u$ quark in proton under bare tensor vertex --- Eq.\eqref{ten-dist-bare} in Fig.\ref{fig-ten-dist}.

Next we explore the dressing effects on tensor charge.
Substituting the quark tensor vertex solution $\Gamma^{\mu \nu}(k; 0)$ obtained from RL truncation scheme or beyond RL scheme, we can get the tensor charge of $u$ quark in the proton for the corresponding truncation scheme.
Substituting the general form of tensor vertex one can obtain
\begin{equation}
\begin{split}
h_{RL}(k^2) &= h_e(k^2) E_{RL}(k^2) + h_f(k^2) F_{RL}(k^2)\\
&\quad + h_h(k^2) H_{RL}(k^2)\, ,
\end{split}
\end{equation}
\begin{equation}
\begin{split}
h_{MU}(k^2) &=  h_e(k^2) E_{MU}(k^2) + h_f(k^2) F_{MU}(k^2)\\
&\quad + h_h(k^2) H_{MU}(k^2)\, ,
\end{split}
\end{equation}
where $h_e(k^2), h_f(k^2), h_h(k^2)$ are the tensor charge distribution functions corresponding to the scalar function in tensor vertex amplitude $E(k^2), F(k^2), H(k^2)$ respectively. The subscribes ``$RL$'' and ``$MU$'' stand for the tensor vertex amplitude solution in RL approximation and tensor vertex amplitude solution in Munczek ansatz. The dressing functions take the following forms:
\begin{subequations}
\begin{align}
h_e(k^2) &= (\frac{g \Lambda^2}{4 \pi})^2 \frac{Z k^2}{(k^2 + m^2)^2 (k^2 + \Lambda^2)^2}\times\\
&\nonumber\quad \big[m^2 + M^2 \frac{(Z k^2)^2}{3} + m M Z k^2\big]\, ,\\
h_f(k^2) &= (\frac{g \Lambda^2}{4 \pi})^2 \frac{Z k^2}{(k^2 + m^2)^2 (k^2 + \Lambda^2)^2}\times\\
&\nonumber\quad \big[-\frac{2}{3}m M^2 (Z k^2)^2 + 2 m k^2\\
&\nonumber\quad\; - m^2 M Z k^2 + M k^2 Z k^2\big]\, ,\\
h_h(k^2) &= (\frac{g \Lambda^2}{4 \pi})^2 \frac{Z k^2}{(k^2 + m^2)^2 (k^2 + \Lambda^2)^2}\times\\
&\nonumber\quad \big[\frac{1}{3}m^2 M^2 (Z k^2)^2 - m M k^2 Z k^2 + (k^2)^2\big]\,.
\end{align}
\end{subequations}
The momentum dependence of tensor charge distributions are shown in Fig.\ref{fig-ten-dist}.

\begin{table}[b]
\centering
\caption{The $u$ quark tensor charge in proton under bare, rainbow-ladder, beyond rainbow-ladder scheme. The error bars comes from varying $\omega$ under the same quark condensate: $\omega\in[0.50,0.54]$ in RL approximation and $\omega\in[0.345,0.365]$ in the case of Munczek ansatz.}
\begin{tabular}{|c|c|c|c|}
\hline
truncation scheme & bare & rainbow-ladder & Munczek\\
\hline
tensor charge $\delta u$ & 0.898   & $0.687^{+0.007}_{-0.006}$  & $0.782^{+0.016}_{-0.013}$\\
\hline
\end{tabular}
\label{tab-ten-com}
\end{table}

By integrating out momentum $k^2$, we get the $u$ quark tensor charge in proton under the three truncation schemes. The obtained results are listed in Table.\ref{tab-ten-com}. The error bars comes from varying $\omega$ under the same quark condensate: $\omega\in[0.50,0.54]$ in RL approximation and $\omega\in[0.345,0.365]$ in the case of Munczek ansatz.  Our results are consistent with the calculations from lattice QCD.

Besides, when considering the dressing effects on tensor vertex, the tensor charges are suppressed than the bare tensor vertex by about $23\%$ in the case of RL approximation. When going beyond the RL approximation,  the tensor charges increases a bit, and the suppression of $u$ quark tensor charge turns out to be about $13\%$ than the result of bare tensor vertex.

Thus in general, the dressing effects suppress the tensor charge significantly. The RL truncation dominates and the effects  beyond RL approximation give corrections --- make the tensor charge increases about $10\%$.

The weakness of this simple picture of nucleon is obvious. It does not contain nonzero value of angular momentum component of scalar amplitude and axial-vector component. Both are important when one charts the large momentum behavior of electromagnetic form factor of nucleon. This deficiency forbids us  to produce down quark tensor charge at the same level. The momentum dependence of quark-diquark amplitude is just the one carried by quark which is  not realistic and could be improved in the future work.

\begin{figure}[h]
\centering
\includegraphics[width=0.5\textwidth]{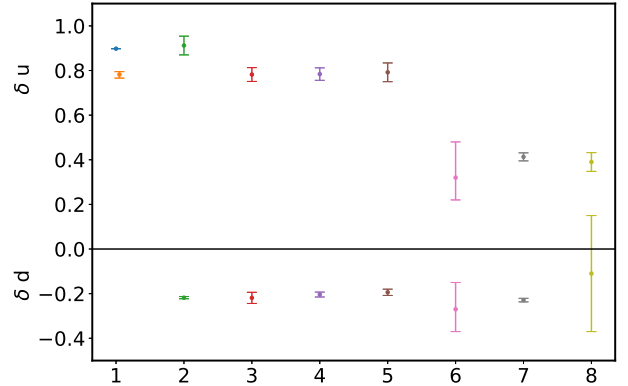}
\caption{(Color online)Some results of quark tensor charges from ours, Faddeev equation, lattice QCD and experiments. The points with label 1 are the results of our calculations with the bare quark tensor vertex (upper blue point) and quark tensor vertex within Munczek ansatz (lower orange point). The green points with label 2 are the results from Faddeev equation \cite{PhysRevD.98.054019}. The points with label 3,4,5 are the results from lattice QCD calculations \cite{PhysRevD.94.054508,Alexandrou:2017qyt,Gupta:2018lvp}. The points with label 6,7,8 are the results extracted from experimental data \cite{PhysRevD.87.094019,Ye:2016prn,PhysRevLett.120.192001}.}
\label{fig-compare-data}
\end{figure}

Finally,  we depict the results of quark tensor charges in proton from ours, Faddeev equation, lattice QCD and experiments in fig. \ref{fig-compare-data}. Quantitatively, our results of quark tensor charges in proton calculated from Munczek ansatz are consistent with the results from Faddeev equation as well as the lattice QCD calculations, but all the results from  theoretical calculations  are quite different from the experimental data analysis, which needs to be confirmed more carefully on both theoretical and experimental side.

%%%%%%%%%%%%%%%%%%%%%%%%%%%%%%%%%%%%%%%%%%%%%%%%%%%%%%%%%%%%%%%%%%%%%%%%%%%%%%%%%%%%%%%%%%%%%%
\section{Summary and Remarks}
\label{sec:sum}
%%%%%%%%%%%%%%%%%%%%%%%%%%%%%%%%%%%%%%%%%%%%%%%%%%%%%%%%%%%%%%%%%%%%%%%%%%%%%%%%%%%%%%%%%%%%%
In this paper, we calculate the quark tensor vertex within the consistent framework of Dyson-Schwinger equation and Bethe-Salpeter equation. We go beyond the rainbow-ladder approximation by utilizing the Munczek ansatz of quark-gluon vertex which satisfies the Ward-Takahashi identity.   We construct the two particle irreducible scattering kernel directly from the functional derivative of self energy. We find the obtained kernel in this way is consistent with that obtained from symmetry constraints .

We then  combine the dressed tensor vertex  with a simple nucleon model with only  quark and scalar diquark components being taken into account.
By comparing the dressed tensor vertex amplitude with the tree level tensor vertex, we show that the dressing effect suppresses the tensor charge by about $23\%$ in RL approximation.  When going beyond the RL approximation, the tensor charge increases a bit compared to that in rainbow-ladder approximation, the suppression becomes only $13\%$ to the tree-level tensor vertex. This means the rainbow-ladder contribution of tensor charge dominates in nucleon and the contribution of beyond RL approximation gives $10\%$ correction in the nucleon tensor charge.

Although we estimate the dress effects of tensor charge in a simple model, this estimation can illustrate the dressing effects of tensor charge in nucleon to large extent. The  obtained $u$-quark tensor charge is consistent with the result from lQCD simulations. On  one hand, the scalar diquark channel dominates the proton in momentum range lower than $1\text{GeV}$, which is the scale of proton; on the other hand, the model parameters follow from the fitting of experimental electromagnetic form factors. As long as the nucleon have these electromagnetic form factor, the results in this model are approximately true. 

\section{Acknowledgement}

We are grateful for constructive suggestions from Muyang Chen, Minghui Ding, Fei Gao, Si-xue Qin, K.R.Montano and C. D. Roberts. Work supported by:
the National Natural Science Foundation of China under contracts No. 11435001, and No. 11775041, the National Key Basic Research Program of China under contract No. 2015CB856900, and 
the Chinese Government's \emph{Thousand Talents Plan for Young Professionals}.

\end{document}